\documentclass{phc-proc4-auth}
\usepackage{graphicx}
\usepackage{amsmath}

\begin{document}
\begin{frontmatter}
\title{The Hubbard model: \protect\\ bosonic excitations and zero-frequency constants}
\author{Adolfo Avella\corauthref{avella}}
\ead{avella@sa.infn.it} \ead[url]{http://scs.sa.infn.it/avella}
\corauth[avella]{Dipartimento di Fisica "E.R. Caianiello"
\protect\\ Unit\`a di Ricerca INFM di Salerno \protect\\
Universit\`{a} degli Studi di
Salerno \protect\\ Via S. Allende, I-84081 Baronissi (SA), Italy \protect\\
Tel.  +39 089 965418 \protect\\ Fax:  +39 089 965275}
\author{Ferdinando Mancini}
\ead{mancini@sa.infn.it} \ead[url]{http://scs.sa.infn.it/mancini}
\address{Dipartimento di Fisica ``E.R. Caianiello'' - Unit\`a INFM di Salerno \protect\\
Universit\`a degli Studi di Salerno, I-84081 Baronissi (SA),
Italy}
\begin{abstract}
A fully self-consistent calculation of the bosonic dynamics of the
Hubbard model is developed within the Composite Operator Method.
From one side we consider a basic set of fermionic composite
operators (Hubbard fields) and calculate the retarded propagators.
On the other side we consider a basic set of bosonic composite
operators (charge, spin and pair) and calculate the causal
propagators. The equations for the Green's functions (GF)
(retarded and causal), studied in the polar approximation, are
coupled and depend on a set of parameters not determined by the
dynamics. First, the pair sector is self-consistently solved
together with the fermionic one and the zero-frequency constants
(ZFC) are calculated not assuming the ergodic value, but fixing
the representation of the GF in such a way to maintain the
constrains required by the algebra of the composite fields. Then,
the scheme to compute the charge and spin sectors, ZFCs included,
is given in terms of the fermionic and pair correlators.
\end{abstract}
\end{frontmatter}

Recently, the Green's function method for composite operators has
been revisited \cite{Mancini:00}. In particular, it has been shown
that the formulation generates an internal self-consistency which
cannot be solved uniquely by the dynamics, but ingredients related
to the microscopic nature of the local operator algebra and to the
macroscopic nature of the external boundary conditions must be
provided. This is not surprising. The properties of composite
operators are not known at priori, they have a microscopic nature
but manifest at level of observation; as a consequence they are
self-consistently determined by the dynamics of the system, by the
algebra and by the boundary conditions.

Roughly, the properties of electronic systems can be classified in
two large classes: single particle properties, described in terms
of fermionic propagators, and response functions, described in
terms of bosonic propagators. These two sectors, fermionic and
bosonic, are not independent, and a fully self-consistent solution
requires that both sectors are simultaneously solved. All the new
theoretical schemes, developed in the last years, show the
importance of the spin and charge correlations in order to
describe the single particle properties of highly correlated
electron systems.

In order to illustrate these ideas, we consider the Hubbard model,
described by the Hamiltonian
\begin{equation}
H=\sum_{\bf i,j}(t_{\bf ij}-\mu \delta_{\bf ij})c^\dagger ({\bf
i},t)c({\bf j},t)+U\sum_{\bf i} n_\uparrow (i) n_\downarrow (i)
\end{equation}
We use the standard notation: $c(i)$, $c^\dagger (i)$ are
annihilation and creation operators of electrons in the spinor
notation; ${\bf i}$ stays for the lattice vector and $i = ({\bf
i},t)$; $\mu$ is the chemical potential; $t_{\bf ij}$ denotes the
transfer integral; $U$ is the screened Coulomb potential;
$n_\sigma (i)=c_\sigma ^\dagger (i)c_\sigma (i)$ is the charge
density of electrons at the site {\bf i} with spin $\sigma$. For a
cubic lattice and by considering only nearest neighbor sites the
hopping matrix has the form $t_{\bf ij}=-2dt\alpha_{\bf ij}$,
where $d$ is the dimension and $\alpha_{ij}$ is the projection
operator
\begin{equation}
\alpha_{\bf ij}={1 \over N}\sum_{\bf k} {e^{{\rm i}{\bf k} \cdot
(R_{\bf i}- R_{\bf j})}}\alpha ({\bf k}) \quad \quad \alpha ({\bf
k})={1 \over d}\sum_{n=1}^d \cos (k_n)
\end{equation}

We choose as fermionic basis
\begin{equation}
\psi (i)=\left(
\begin{matrix}
\xi (i)\\
\eta (i)
\end{matrix}
\right)
\end{equation}
where $\xi (i)=[1-n(i)]c(i)$ and $\eta(i)=n(i)c(i)$ are the
Hubbard operators, and $ n(i)=\sum_\sigma n_\sigma(i)$. In the
two-pole approximation \cite{Avella:98} the retarded GF $G(i,j)=
\langle R[\psi (i)\psi^\dagger (j)] \rangle $ satisfies the
equation
\begin{equation}
[\omega -\varepsilon ({\bf k})]G(k,\omega )=I({\bf k})
\end{equation}
where $I({\bf k})=F.T. \langle \{ \psi ({\bf i},t),\psi ^\dagger
({\bf j},t)\} \rangle $ and $ \varepsilon ({\bf k})=F.T. \langle
\{ {\rm i}{{\partial \psi ({\bf i},t)} \over {\partial t}},\psi
^\dagger ({\bf j},t)\} \rangle I^{-1}({\bf k})$; the symbol $F.T.$
denotes the Fourier transform. In the paramagnetic phase the
energy matrix $\varepsilon ({\bf k})$ depends on the following set
of internal parameters: $\mu$, $\Delta = \langle \xi ^\alpha
(i)\xi ^\dagger (i) \rangle- \langle \eta ^\alpha (i)\eta ^\dagger
(i) \rangle$, $p= \langle n_\mu ^\alpha (i)n_\mu (i) \rangle /4-
\langle [c_\uparrow (i)c_\downarrow (i)]^\alpha c_\downarrow
^\dagger (i)c_\uparrow ^\dagger (i) \rangle$, which must be
self-consistently determined. Given an operator $\zeta(i)$, we are
using the notation $\zeta ^\alpha (i)=\sum_{\bf j} \alpha _{\bf
ij} \zeta ({\bf j},t)$. The operator $n_\mu (i)=c^\dagger (i)
\sigma _\mu c(i)$ [$\sigma _\mu =({\bf 1},\mathbf{\sigma} )$,
$\sigma$ are the Pauli matrices] is the charge ($\mu = 0$) and
spin ($\mu = 1,2,3$) density operator. The local algebra satisfied
by the fermionic field (3) imposes the constraint $\langle \xi
(i)\eta ^\dagger (i) \rangle =0$: this equation allows us to solve
self-consistently the fermionic sector. However, it is worth
noticing that the presence of the parameter $p$ directly relates
the fermionic sector to bosonic sectors.

We consider then the composite bosonic field
\begin{equation}
N^{(\mu )}(i)=\left(
\begin{matrix}
n_\mu (i)\\
\rho_\mu (i)
\end{matrix}
\right)
\end{equation}
where $\rho_\mu(i)=c^\dagger (i)\sigma _\mu c^\alpha (i)-c^{\alpha
\dagger} (i)\sigma _\mu c(i)$. In the two-pole approximation
\cite{Avella:03} the causal GF $G^{(\mu )}(i,j)= \langle T[N^{(\mu
)}(i)N^{(\mu )\dagger} (j)] \rangle$ satisfies the equation
\begin{equation}
[\omega -\varepsilon ^{(\mu )}({\bf k})]G^{(\mu )}({\bf k},\omega
)=I^{(\mu )}({\bf k})
\end{equation}
where $I^{(\mu)}({\bf k})=F.T. \langle [N^{(\mu)}({\bf
i},t),N^{(\mu)\dagger} ({\bf j},t)] \rangle$ and $\varepsilon
^{(\mu )}({\bf k})=F.T. \langle [{\rm i}{{\partial N^{(\mu )}({\bf
i},t)} \over {\partial t}},N^{(\mu )\dagger} ({\bf j},t)] \rangle
[I^{(\mu )}({\bf k})]^{-1}$. In the one-dimensional case (we
consider the 1D case just for the sake of simplicity) the energy
matrix $\varepsilon ^{(\mu )}({\bf k})$ depends on the following
set of internal parameters: (i) fermionic parameters: $C^\alpha =
\langle c^\alpha (i)c^\dagger (i) \rangle$, $C^\eta = \langle
c^\eta (i)c^\dagger (i) \rangle $, $C^\lambda = \langle c^\lambda
(i)c^\dagger (i) \rangle$, $E= \langle \eta (i)\eta ^\dagger (i)
\rangle$, $E^\eta = \langle \eta ^\eta (i)c^\dagger (i) \rangle $,
where $\eta _{ij}$, $\lambda_{ij}$ are the projection operators on
the second and third nearest neighbors, respectively; (ii) bosonic
parameters: $a_\mu$, $b_\mu$ and $c_\mu $, whose explicit
expressions, although for the 2D case, are reported in
Ref.~\cite{Avella:03}; (iii) zero-frequency functions (ZFF)
$\Gamma^{(\mu )}({\bf i,j})$ (see Ref.~\cite{Mancini:00}). Due to
the hydrodynamic constraints, two bosonic parameters should be
determined as $b_\mu =a_\mu +n-2(E-E^\eta )$ and $c_\mu =a_\mu
-n+2(E-E^\eta )$. The parameter $a_\mu$ instead can be determined
by means of the local algebra constrain $\langle n_\mu (i)n_\mu
(i) \rangle = \langle n \rangle +2(2\delta _{\mu ,0}-1)D$, where
$D=\langle n \rangle/2-E$ is the double occupancy. The ZFF are
left undetermined.

\begin{figure}[tb]
\begin{center}
\includegraphics[height=7cm,angle=270]{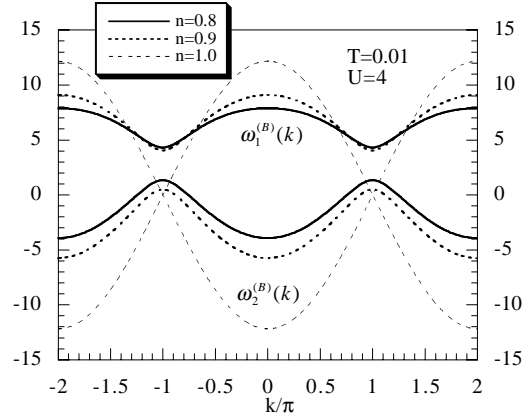}
\end{center}
\caption{Pair energy spectra $\omega _n^{(B)}(k)$ as a function of
the momentum $k$ for Coulomb repulsion $U=4$, temperature $T=0.01$
and filling $n=0.8$, $0.9$ and $1$.} \label{Fig1}
\end{figure}

\begin{figure}[tb]
\begin{center}
\includegraphics[height=7cm,angle=270]{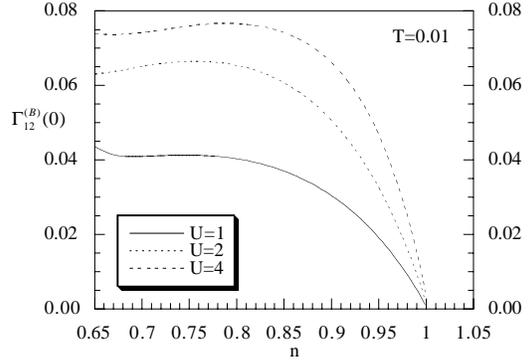}
\end{center}
\caption{Pair zero-frequency constant $\Gamma_{12}^{(B)}({\bf 0})$
as a function of the filling $n$ for temperature $T=0.01$ and
Coulomb repulsion $U=1$, $2$ and $4$.} \label{Fig2}
\end{figure}

According to this, we need another composite bosonic field
\begin{equation}
B(i)=\left(
\begin{matrix}
h(i)\\ f(i)
\end{matrix}
\right) \notag
\end{equation}
where $h(i)=c_\uparrow (i)c_\downarrow (i)$ and $f(i)=c_\uparrow
^\alpha (i)c_\downarrow (i)+c_\uparrow (i)c_\downarrow ^\alpha
(i)$. In the two-pole approximation the causal GF $G^{(B)}(i,j)=
\langle T[B(i)B^\dagger (j)] \rangle$ satisfies the equation
\begin{equation}
[\omega -\varepsilon^{(B)}(k)]G^{(B)}(k,\omega )=I^{(B)}(k) \notag
\end{equation}
where $I^{(B)}(k)=F.T.\langle[B({\bf i},t),B^\dagger ({\bf
j},t)]\rangle$ and $\varepsilon ^{(B)}(k)=F.T. \langle [{\rm
i}{{\partial B({\bf i},t)} \over {\partial t}},B^\dagger ({\bf
j},t)] \rangle [I^{(B)}(k)]^{-1}$. The energy matrix $\varepsilon
^{(B)}(k)$ depends on the following set of internal parameters:
(i) fermionic parameters: $C^\alpha$, $C^\eta$, $C^\lambda$, $E$,
$E^\eta$; (ii) bosonic parameters: $u$, $v$ and $w$, whose lengthy
expressions are not reported here for the sake of brevity; (iii)
ZFF $\Gamma^{(B)}({\bf i,j})$. The condition that the pair energy
spectra $\omega _n^{(B)}(k)$ are finite and the local algebra
constrain $\langle h(i)h^\dagger (i)\rangle=1-\langle n \rangle
+D$ completely determine the parameters $u$, $v$ and $w$. The
algebra constraint $\langle h(i)f^\dagger (i) \rangle =2\langle
\xi ^\alpha (i)c^\dagger (i) \rangle $ will be used to compute the
ZFC $\Gamma_{12}^{(B)}({\bf 0})$. Then, the pair sector can be
immediately solved once the solution for the fermionic sector has
been found. The results of this scheme are shown in
Figs.~\ref{Fig1} and \ref{Fig2}, where the pair energy spectra and
$\Gamma_{12}^{(B)}({\bf 0})$ are shown, respectively. It is worth
noticing that, in the present scheme, the pair dynamics is ergodic
only at half-filling.

Now, once we have solved the fermionic and pair sectors we can
come back to the charge-spin one. We need to compute the six ZFC:
$\Gamma _{11}^{(0)}(0)$, $\Gamma _{11}^{(0)}(a)$,
$\Gamma_{11}^{(3)}(0)$, $\Gamma_{11}^{(3)}(a)$, $\Gamma
_{22}^{(0)}(0)$, $\Gamma_{22}^{(3)}(0)$. They can be fixed by
means of as many algebra constrains coming from the expressions of
the following correlators: $\langle n_\mu (i)n_\mu
^\alpha(i)\rangle$, $\langle \rho_\mu (i)\rho_\mu
^\alpha(i)\rangle$, $\langle h^\alpha (i)h^\dagger (i)\rangle$ and
$\langle f(i)f^\dagger (i)\rangle$. The explicit expressions of
the constraints are quite lengthy and will be given elsewhere.

In conclusion, we have reported a fully self-consistent scheme of
calculations for both the fermionic and the three (spin, charge
and pair) bosonic sectors of the Hubbard model. It is worth
noticing that, within this scheme, the ZFC of the spin and charge
sectors, which could assume, at least in principle, not ergodic
values as the pair one does, can be self-consistently computed and
give invaluable information regarding the dynamics of the
corresponding operators.


\end{document}